\documentclass{kluwer}    
\usepackage{psfig}

\newdisplay{guess}{Conjecture}
\newcommand{\msun}{\mbox{$M_{\odot}$}}

\newcommand{\lsun}{\mbox{$L_{\odot}$}}

\newcommand{\teff}{\mbox{$T_{\rm eff}$}}
\newcommand{\Teff}{\mbox{$T_{\rm eff}$}}

\newcommand{\ha}{\mbox{H$\alpha$}}
\newcommand{\mdot}{\mbox{$\dot{M}$}}
\newcommand{\Mdot}{\mbox{$\dot{M}$}}

\newcommand{\msunyr}{\mbox{$M_{\odot} {\rm yr}^{-1}$}}

\newcommand{\bb}{\bibitem[]{}}

\begin{document}                                                                                   
\begin{article}
\begin{opening}         
\title{Mass loss predictions for Subdwarf B stars} 
\author{Jorick S. \surname{Vink}}  
\runningauthor{Jorick Vink}
\runningtitle{Mass loss in Subdwarf B stars}
\institute{Imperial College London, Blackett Lab, London, SW7 2BZ, UK}

\begin{abstract}
We present the results of Monte Carlo mass-loss computations for hot low-mass stars,  
specifically for Subdwarf B (SdB) stars. It is shown that the mass-loss rates
{\it on} the Horizontal Branch (HB) computed from radiative line-driven wind models
are not high enough to create SdB stars. 
We argue, however, that mass loss plays a role in the chemical 
abundance patterns observed both in field SdB stars, as well as 
in cluster HB stars. 
The derived mass loss recipe for these (extremely) hot HB stars 
may also be applied to other groups of hot low-mass stars, such as 
post-HB (AGB-manqu{\'e}, UV-bright) stars, over a range in effective temperatures 
between $\simeq$ 10\,000 -- 50\,000 Kelvin. Finally, we present preliminary 
spectral synthesis on the more luminous SdB stars for which emission cores in \ha\ have 
been detected (Heber et al. 2002). We find that these line profiles can indeed 
be interpreted as the presence for a stellar wind with mass-loss of the order 
of $10^{-11}$ \msunyr.
\end{abstract}

\end{opening}           

\section{Introduction}  

Extreme Horizontal Branch (EHB) stars are extremely hot (\teff\ $\ge$ 30 000 K) HB stars of $\simeq$ 0.5 \msun, 
consisting of a helium core and a very thin layer of hydrogen ($\le$ 0.02 \msun) on top (see e.g. Dorman et al. 1993). 
This hydrogen layer is so thin that EHB stars cannot undergo helium shell burning, which prevents them from ascending 
the Asymptotic Giant Branch (AGB). Instead, they are believed to directly evolve towards the white dwarf 
cooling track as ``failed'' AGB (or AGB manqu{\'e} stars). 
Subdwarf B (SdB) star is the other well-known name. Although the ``EHB'' nomenclature is 
mostly found in the context of globular cluster HB stars, the SdB term is generally used to describe 
the objects in the field. The evolutionary status of EHB/SdB stars is as yet a mystery, so the correct 
evolutionary scenario has yet to be identified.
One of the key observational constraints that can help distinguish between the different proposed 
evolutionary channels (see e.g. Sweigart, Podsiadlowski, Jeffery, and others in these proceedings) 
involves the stellar abundances in SdB stars.  
Interestingly, SdB stars are characterised by strong chemical abundance anomalies: (i) helium is severely 
depleted: underabundances by factors of hundreds to thousands 
are not uncommon, while (ii) metals (such as carbon, oxygen and nitrogen) show similarly unexplained wide spreads. 
Gravitational settling and radiative levitation have been suggested 
as the most natural cause for the striking abundance anomalies, but in recent years the necessity for 
a stellar wind has become clear: without a wind, helium ``sinks in'' (due to gravity) in too short 
a timescale compared to the evolutionary lifetime of an EHB object (Fontaine \& Chayer 
1997, Unglaub \& Bues 2001). 
Current atmospheric diffusion calculations treat SdB mass loss as a free parameter, but these diffusion 
computations can be made more robust if reliable mass-loss rates are provided. 
One way to reach such goal would be via observational results, however sensitive observational 
techniques are yet lacking (but see Sect.~\ref{s_outlook} for an outlook). One therefore needs to rely 
on the theory of radiative line-driven winds. 
In this context, Vink \& Cassisi (2002) have recently computed radiation-driven wind models for HB stars, 
and discussed stellar winds in the context of the ``zoo'' of problems governing HB morphology in globular 
clusters occurring at \teff\ $\simeq$ 11~000 K. They have argued that stellar winds can resolve these 
issues in a natural way: a stellar wind that is set-up by the increase in metallicity (due to radiative 
levitation) for HB stars with \teff\ $\ge$ 10 000 K.

\section{Motivation for accurate mass-loss rates}
\label{s_mot}

We identify five reasons as to why one should 
be concerned about the mass-loss rates for (E)HB/SdB stars:

\begin{enumerate}

\item{} Does mass loss on the HB affect the evolution of HB stars? 
This question was recently posed by Yong et al. (2000). The authors hypothesised that
a stellar wind {\it on} the HB could strip enough matter off the stellar core so that 
a blue HB star naturally evolves into an EHB/SdB star. If this scenario were correct, the problem 
of the existence of SdB stars would be solved. 
Since accurate mass-loss rates for HB stars were lacking, Yong et al. had the freedom to apply 
mass-loss rates on order of $10^{-9}$ \msunyr, which appear {\it a posteriori} on the high side. 
Nonetheless, even mass-loss rates a few magnitudes lower than this number, approximately 
$10^{-11}$ \msunyr, might still {\it directly} influence stellar evolution models (see Vink \& 
Cassisi 2002 for details).

\item{} Do stellar winds affect the observed spectra of (E)HB stars?
For massive O stars, it is well-known that the neglect of winds can lead 
to severe errors in the spectroscopic mass determination compared to evolutionary models: 
``the mass discrepancy''(see e.g. Herrero et al. 1992). 
Since there also appears to be ``log $g$'' problem (i.e. a mass discrepancy) in Blue HB 
stars (Moehler et al. 1995, 2000), one may wonder whether there could be potential 
problems using hydrostatic model atmospheres 
for these objects also (see Vink \& Cassisi 2002 for results on the gravity sensitive H$\gamma$ line).

\item{} Does mass loss affect the angular momentum distribution in a rotating (E)HB star?
Vink \& Cassisi (2002) have shown that the difference in predicted mass-loss rate 
between 'cool' ($\teff <$ 10~000 K) and 'warm' ($\teff >$ 10~000 K) HB stars is a factor 
of 100, because of an increased metal abundance 
above this temperature. Since this coincides exactly with an unexplained drop in the rotational velocities 
above this temperature (Behr 1999, 2000), it is tempting to attribute this to the removal 
of angular momentum by a stellar wind (see also Sweigart 2000).

\item{} Does mass loss affect SdB abundances? Computations performed by both Fontaine \& Chayer (1997), 
as well as Unglaub \& Bues (2001) have shown a need for the presence of stellar mass loss 
in the range of $10^{-14}$ $\le$ \mdot\ (\msunyr) $\le$ $10^{-12}$.

\item{} Last, but not least: how does radiative driving behave over the Hertzsprung-Russell Diagram? 
In the following, we describe the Monte Carlo technique simulating photon 
transfer through a unified stellar atmosphere (including a wind). The technique has 
been very successful in (i) explaining the 
bi-stability jump in B supergiants (Vink et al. 1999), (ii) the mass-loss rates of massive O 
stars (Vink et al. 2000), and (iii) the capricious mass loss behaviour of Luminous Blue 
Variables (Vink \& de Koter 2002). By extending the computational method to different domains 
of the Hertzsprung-Russell Diagram, our understanding of radiative line driving can be further enhanced.

\end{enumerate}

\section{The Monte Carlo Method}
\label{s_method}

The description of the radiative wind driving 
with our method is based on a Monte Carlo method 
that was first introduced by Abbott \& Lucy (1985).
This approach naturally accounts for multi-line transfer 
simulating photon-interactions with different metal ions, 
both lines as well as continua, while the photons attempt
to escape the gravitational well of the star.
In the models used here, the ionisation and excitation 
for the dominant ionic species are properly computed 
using the non-LTE unified Improved Sobolev Approximation code 
({\sc isa-wind}; de Koter et al. 1993, 1997), which treats  
the photosphere and wind in a unified manner.
The chemical species that are correctly calculated are 
H, He, C, N, O, and Si. The iron-group elements however 
are treated only approximately using a generalised version 
of the ``modified nebular approximation'' developed by Lucy (1987).

One of the main assumptions implicit in our method is that 
the plasma behaves as a single fluid. As 
long as a large number of collisions between the
accelerating (C,N,O, and Fe-group) and 
non-accelerating (H and He) particles ensures  
a strong coupling, one can safely 
treat the wind as a single fluid. Test calculations 
performed in Vink \& Cassisi (2002) have shown that  
Coulomb coupling is fulfilled (albeit only within an 
order of magnitude).
Lastly, it remains yet to be seen if the use 
of the Sobolev approximation is valid for weaker 
winds (see Owocki \& Puls 1999). 

\section{Mass loss rates}
\label{s_results}

The mass loss rates for EHB/SdB stars are readily obtainable from the 
mass loss recipe presented by Vink \& Cassisi (2002):

\begin{eqnarray}
{\rm log}~\dot{M} & = &~-11.70~(\pm 0.08) \nonumber \\
                  & &~+~1.07~(\pm 0.32)~{\rm log} (\teff/20000) \nonumber\\
                  & &~+~2.13~(\pm 0.09)~({\rm log}L_* - 1.5) \nonumber\\
                  & &~-~1.09~(\pm 0.05)~{\rm log}(M_*/0.5) \nonumber\\
                  & &~+~0.97~(\pm 0.04)~{\rm log}(Z_*) \nonumber\\
                  \nonumber\\
                  & &~{\rm derived~for:} \nonumber\\
                  & &              \,12\,500 \le \teff \le 35\,000\,{\rm K} \nonumber\\
                  & &                  \,1.3 \le {\rm log}L_*  \le 1.7 \nonumber\\
                  & &                   \,0.5 \le {M_*} \le 0.7 \nonumber\\
                  & &                   \,0.1 \le {Z_*} \le 10 \nonumber\\
\label{eq_HBfit}
\end{eqnarray}
where \Teff\ is in Kelvin and $L_*$, $Z_*$, and $M_{*}$ are all given 
in solar units. One of the outcomes is that bi-stability jumps (due to iron recombinations) in the 
mass-loss rate are absent on the HB, as the winds are ``thin'' compared to those of 
OB supergiants. The many weak lines of the element iron (Fe) are not as dominant in 
setting the mass-loss rate, as they are in denser winds (such as OB supergiants). Instead, 
lighter metals, such as carbon, nitrogen, and oxygen, which have fewer, but stronger lines  
are capable of playing an important role in determining the mass loss in thinner winds, 
such as low metallicity O star winds (see Vink et al. 2001 for a fuller discussion). 
Note that for helium rich SdB 
stars (He-SdB, blue hook stars), which {\it may} be explained by the helium mixing scenario 
(Castellani \& Castellani 1993, Brown et al. 2001, Cassisi et al. 2003), carbon is 
predicted to be enriched, which 
may therefore increase the expected mass loss to values {\it above} those following from 
the Vink \& Cassisi recipe.
    
Nonetheless, Eq.~\ref{eq_HBfit} may be applied 
to hot, low-mass stars, of the types: (E)HB, sdB, sdOB, post-HB, AGB-manqu{\'e}, 
UV-bright stars, in a range of \teff\ between 10~000 and 50\,000 K. Central Stars of 
Planetary Nebulae (CSPN) are more resemblant to massive O stars, and 
therefore the OB star recipe of Vink et al. (2001) is recommended for these objects.
Computer routines (IDL) for both mass loss recipes are available on the web 
\footnote{http://astro.ic.ac.uk/$\sim$jvink/}. 

The important points regarding the actual mass-loss values are: (i)  
the computed rates do not support the evolutionary channel of producing 
SdB stars by mass loss on the HB, but (ii) they are in the appropriate 
range of the values deemed necessary in the diffusion computations of Unglaub \& 
Bues (2001), i.e. the observed chemical patterns can only be explained 
if mass-loss rates are in the range $10^{-14}$ $\le$ \Mdot\ (\msunyr) $\le$ $10^{-12}$. 
Higher rates would prevent the effect of diffusion, whereas for 
lower rates helium would sink in too short time scales.

We conclude that our mass-loss rates are in the right ballpark as far 
as consistency with diffusion calculations is concerned. Nonetheless, what one 
finally seeks is confirmation by observations. 

\section{Outlook on observational tests}
\label{s_outlook}

\begin{figure} 
\centerline{\psfig{file=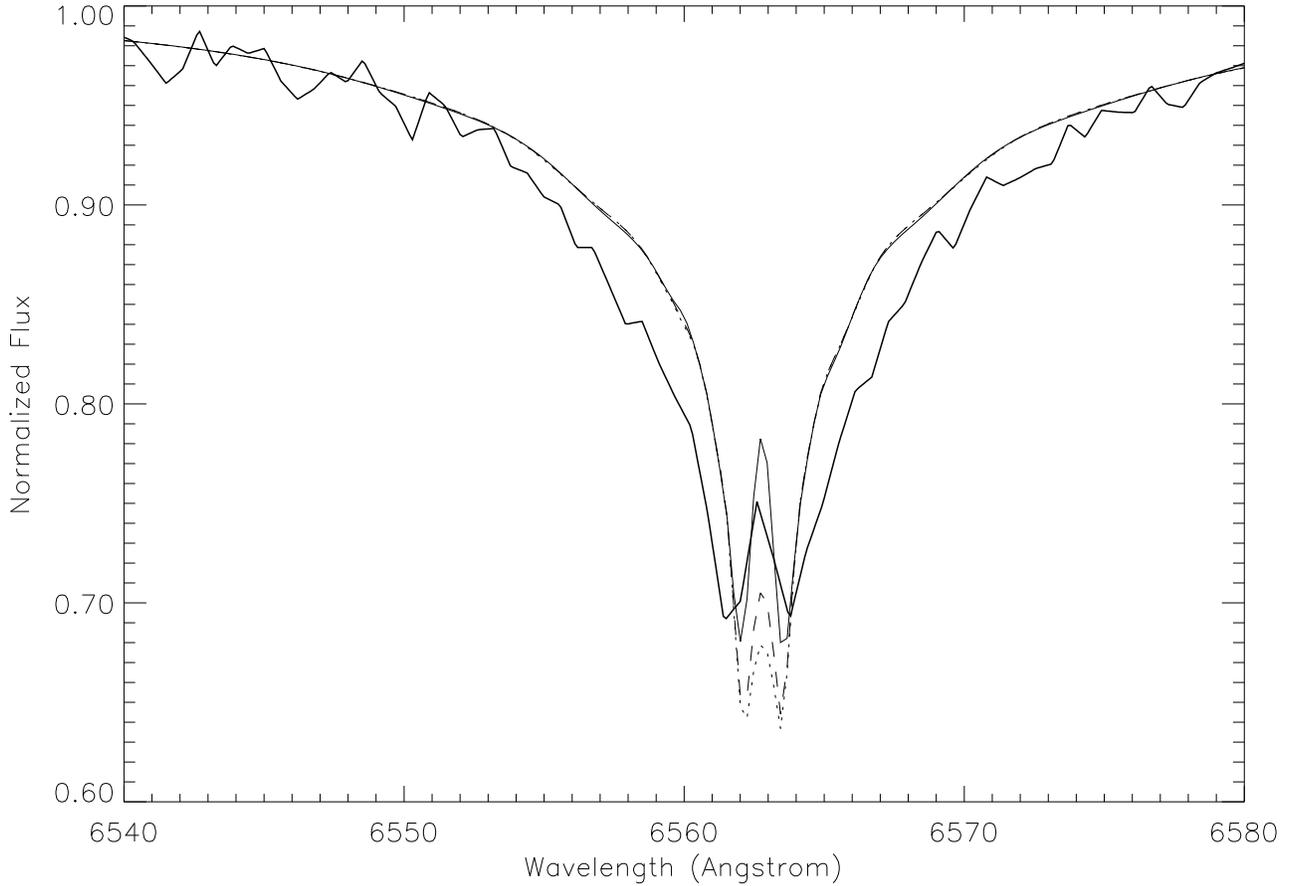}}
\caption{The thick solid line represents the observed \ha\ line of PG 1000$+$408 -- 
a typical \ha\ anomaly presented by Heber et al. (2002). 
The thin solid, dashed, and dotted lines represent spectral models for mass loss rates 
of log \mdot\ (\msunyr) = -10.5, -11.0, and -11.5 respectively. These models are 
calculated for a mass of $M = 0.5 \msun$,  log$(L/\lsun) = 1.51$, and solar abundance.
Note that these are {\it not} intended to be fits to the observed profile (see text
for a discussion).}
\label{f_comp}
\end{figure}

The most sensitive diagnostic for hot star winds is generally accepted to 
be the metal resonance lines in the ultraviolet part of the electromagnetic 
spectrum (e.g. Lamers \& Morton 1976). However, similar data for SdB stars
is currently not available. Another potential diagnostic is \ha\ emission, but 
SdB stars generally show this line in absorption. One may wonder whether there is any chance 
for detecting wind emission in these low luminosity (and thus low wind density) objects at all.
The positive aspect is that SdB stars have small radii, and thus the more relevant 
parameter for detecting mass loss in these objects, does not concern the mass-loss rate, but
rather the mass flux. Based on the results of the Monte Carlo presented above, one may 
expect similar mass fluxes is SdB stars as in the more massive late-O-type/early-B-type 
main sequence stars. Interestingly, Heber et al. (2002) detected anomalous \ha\ lines 
in four SdB stars, and proposed that these anomalies {\it could} be 
the signature of a weak stellar wind.  

To test this hypothesis, we have made a preliminary spectral synthesis of \ha\ using the 
{\sc isa-wind} code (de Koter et al. 1993) for extended atmospheres, and first results 
are presented in Fig.~\ref{f_comp}. The thick solid line shows the observed \ha\ profile of 
PG 1000$+$408, a typical \ha\ anomaly presented by Heber et al. (2002): a small, but clear, 
emission at line centre. Interestingly, our
spectral synthesis (the three thin lines in Fig.~\ref{f_comp}) shows 
similar behaviour of the line core. Note that the 
three thin lines (solid, dashed and dotted) are {\it not} fits to the observed profile, i.e. 
we would need to (i) apply rotational and instrumental convolution to our predicted profiles, and (ii)
synthesise the entire blue and red parts of the spectrum, as to find consistent wind and photospheric 
parameters (most notably log $g$).  

The stellar parameters for PG 1000$+$408 (as determined from {\it hydrostatic} stellar atmosphere analysis) 
are: \teff\ = 36~000 K, log (L/\lsun) = 1.51, assuming a stellar mass of M = 0.5 \msun. 
Our strategy was the following. Adapting the hydrostatic stellar parameters, and assuming solar abundances, 
we made a mass loss prediction using the Vink \& Cassisi mass-loss recipe. The mass loss that is expected 
for PG 1000$+$408 is log $\mdot (\msunyr) = -11.40$ . Since mass loss depends almost linearly 
on metal abundance (as log \mdot\ $\propto$ $Z^{0.97}$; Eq. 1) and the abundances of the objects 
are not known, the actual mass loss could be somewhat different. Therefore, we 
show the effect of varying the mass-loss rate on the predicted line profiles of \ha\ in Fig.~\ref{f_comp}.
The thin solid, dashed, and dotted lines represent mass loss rates of log \mdot\ (\msunyr) = -10.5, -11.0, and -11.5 
respectively.
It is reassuring to find that the strength of the central line emission, now interpreted as wind emission, 
is a function of wind density. Although this is a preliminary analyses, the results are encouraging, 
and indicate that the future of SdB mass loss determinations using optical lines may not at all 
be as hopeless as has previously been thought. 

\section{Summary \& Conclusions}

We have made mass loss predictions for SdB stars using a Monte Carlo method simulating multi-line
interactions in a moving stellar atmosphere. The results show that HB winds are too weak to directly influence 
stellar evolution, and cannot create SdB stars as the sole effect. 
However, the puzzling chemical abundance patterns found in both field and cluster EHB stars are interpreted 
as due to the competing effects of gravitational settling and radiative levitation in the presence of a stellar wind. We have 
succeeded in providing a set of constraints on the wind strengths of EHB/SdB stars, and in turn, these predictions appear to 
meet the constraints set by atmospheric diffusion calculations, as well as the first observational 
tests. The future for mass loss in Subdwarf B stars looks bright!

\acknowledgements
I wish to thank Santi Cassisi for his encouragement and collaboration.

\end{article}
\end{document}